\documentclass[%
    aps,
    prd,
    reprint,
    10pt,
    superscriptaddress,
    amsfonts,
    amssymb,
    amsmath,
    preprintnumbers,
    noshowpacs,
    tightenlines,
    floats,
    notitlepage,
    final,
    letterpaper,
    oneside,
    twocolumn,
    nofootinbib,
    longbibliography,
    ]{revtex4-2}
\usepackage[a4paper, margin=0.8in]{geometry}
\usepackage[table,svgnames,dvipsnames]{xcolor}
\usepackage[colorlinks=true,citecolor=blue,linkcolor=black,urlcolor=blue, backref=false,pdfborder={0 0 0}]{hyperref}
\usepackage[normalem]{ulem}
\usepackage{aas_macros}
\usepackage{graphicx}
\usepackage{dcolumn}
\usepackage{bm}
\usepackage[mathlines]{lineno}
\usepackage{orcidlink}
\usepackage{cases}
\usepackage{booktabs}
\usepackage{comment}
\usepackage{multirow}
\usepackage{siunitx}
\usepackage{booktabs}
\graphicspath{{figs/}}
\usepackage{float}

\usepackage{geometry}

\usepackage{aas_macros}
\usepackage[dvipsnames]{xcolor}
\usepackage{subcaption}
\usepackage{soul}

\def\be{\begin{equation}}
\def\ee{\end{equation}}

\def\ba#1\ea{\begin{align*}#1\end{align*}}

\newcommand{\code}[1]{{\texttt{#1}}}
\renewcommand{\emph}[1]{\textit{#1}}

\newcommand{\refeq}[1]{Eq.~(\ref{eq:#1})}

\newcommand{\reffig}[1]{Fig.~\ref{fig:#1}} 
\newcommand{\reffigs}[2]{Fig.~\ref{fig:#1}--\ref{fig:#2}}

\newcommand{\refsec}[1]{Sec.~\ref{sec:#1}}

\newcommand{\Om}{$\Omega_{\rm M}$}

\newcommand{\pairvel}{$v_{12}(r,a)$}
\newcommand{\lcdm}{$\mathrm{\Lambda CDM}$}
\newcommand{\vot}{$v_{12}$}  

\newcommand{\Mpch}{$h^{-1}\,\mbox{Mpc}$}

\newcommand{\kpchInv}{$\mbox{kpc}/h$}

\newcommand{\elephant}{\textsc{elephant}}

\begin{document}
 

\title{Dynamics of pairwise motions in the fully non-linear regime in LCDM and Modified Gravity cosmologies}

\author{Mariana Jaber\orcidlink{0000-0001-7507-9516}}\email{jaber@cft.edu.pl}
\affiliation{Center for Theoretical Physics, Polish Academy of Sciences, Al. Lotnik\'ow 32/46, 02-668 Warsaw, Poland}

\author{Wojciech A. Hellwing\orcidlink{0000-0003-4634-4442}}
\affiliation{Center for Theoretical Physics, Polish Academy of Sciences, Al. Lotnik\'ow 32/46, 02-668 Warsaw, Poland}

\author{Jorge E. Garc\'ia-Farieta\orcidlink{0000-0001-6667-5471}}
\affiliation{Instituto de Astrof\'{\i}sica de Canarias, s/n, E-38205, La Laguna, Tenerife, Spain}
\affiliation{Departamento de Astrof\'{\i}sica, Universidad de La Laguna, E-38206, La Laguna, Tenerife, Spain}

\author{Suhani Gupta\orcidlink{0000-0003-0955-6761}}
\affiliation{Center for Theoretical Physics, Polish Academy of Sciences, Al. Lotnik\'ow 32/46, 02-668 Warsaw, Poland}

\author{Maciej Bilicki\orcidlink{0000-0002-3910-5809}}
\affiliation{Center for Theoretical Physics, Polish Academy of Sciences, Al. Lotnik\'ow 32/46, 02-668 Warsaw, Poland}


\date{Received 1 December 2023; accepted 16 May 2024; published 14 June 2024}

\begin{abstract}
In contrast to our understanding  of density field tracers, the modeling of direct statistics pertaining to the cosmic velocity field remains open to significant opportunities for improvement. The lack of accurate modeling for the non-linear domain of pairwise velocities restricts our capacity to fully exploit the information encoded in this observable. 
We present a robust approach for modelling the mean infall velocities, \pairvel{},  with broad applicability spanning sub-megaparsec scales and cosmologies extending beyond the standard \lcdm{} paradigm. 
Our approach involves solving the full pair-conservation equation using accurate non-linear power spectrum descriptions. To assess the robustness of our model, we extend it to cosmologies beyond the standard \lcdm{}, in particular, the Hu-Sawicki $f(R)$-gravity and Dvali-Gabadadze-Porrati (DGP) modified gravity models. 
Remarkably, our predictions for pairwise velocities of dark matter particles at kilo-parsec scales exhibit excellent agreement with N-body simulations throughout the entire dynamical range ($0.1 \lesssim \xi \lesssim 1000$, or $r\geq0.4$\Mpch{}). 
Furthermore we show that different gravity models leave distinct signatures in the shape and dynamics of the mean pairwise velocities, providing a potent test of cosmological gravity laws.
\\
\\
(Version accepted for publication in Phys. Rev. D.)
\\
\href{https://doi.org/10.1103/PhysRevD.109.123528}{DOI: 10.1103/PhysRevD.109.123528}

\end{abstract}
\maketitle


\section{\label{sec:intro}Introduction}

Pairwise velocities and the  large-scale velocity fields serve as crucial statistics for large-scale structure (LSS), offering valuable insights that complement the widely utilized density-based measurements.
However, unlike our understanding of cosmic density,  the modelling and understating of direct statistics of the cosmic velocity field still have considerable room for improvement.
This deficiency is one of the main reasons why the rich information encoded in the velocity field
and its potential as a cosmological probe is still largely untapped.

Various approaches have been employed to model the non-linear regime of structure formation in the density field. These include the standard perturbation theory (SPT) (see \cite{2004PhRvD..70h3007S, 2010PhRvD..82f3522T, 2002PhR...367....1B, 2012JCAP...11..029G, 2006PhRvD..73f3519C}), regularized perturbation theory \citep{2014ascl.soft04012T}, fitting formulae like \code{Halofit} and \code{HMCODE} (see  \cite{2019MNRAS.486.1448S, 2012ApJ...761..152T,2015MNRAS.454.1958M}), and  emulator approaches (see eg \cite{2014ApJ...780..111H, 2019PhRvD.100l3540W, 2021MNRAS.505.2840E,Emul2022JCAP}).

As we delve into the  small-scale features of the Universe, it becomes increasingly vital to harness the complementary information embedded in peculiar velocity fields. Pairwise velocities, measuring the typical relative velocity between objects at specific separations, contain information  about the peculiar motions induced by gravity.
As such the information encoded in the moments of the pairwise velocity distribution may be considered as crucial for modelling and  deciphering the non-linear regime. 

The velocity field data has been already effectively utilized in various applications, including redshift-space distortions (RSD) growth-rate estimates (for instance \cite{2012MNRAS.423.3430B, 2015MNRAS.449..848H}), bulk-flow measures (\cite{2009MNRAS.392..743W}), cosmic web analysis (see \cite{2012MNRAS.425.2049H}), measuring the velocity power spectrum (e.g. \cite{2017MNRAS.471.3135H}), or in field reconstructions (see \cite{2023MNRAS.522.5291G, 2021MNRAS.507.1557L} or \cite{2023A&A...670L..15C} where the authors reconstruct the density and velocity fields using \cite{2023ApJ...944...94T}, the most recent update to the Cosmic flows catalog \cite{2012MNRAS.420..447T}).

In the late 1970s, a potent method for modeling mean pairwise velocities (MPV) emerged \citep{1971phco.book.Peebles, 1976ApJ.Peebles, 1977ApJS.Davis.Peebles, 1980lssu.book}. This approach led to the application of the Bogoliubov-Born-Green-Kirkwood-Yvon (BBGKY) hierarchy of equations, which describe the dynamic evolution of self-gravitating particle systems, for modelling of the large-scale cosmic velocity fields. 
Within this hierarchy, the first moment is known as the pair-conservation equation \citep{1976ApJ.Peebles, 1977ApJS.Davis.Peebles}, represented as:
\begin{equation}
    \label{eq:conserv}
    \frac{\partial\xi}{\partial t} + \frac{1}{x^2a}\frac{\partial}{\partial x}[x^2(1+\xi)v]=0.
\end{equation}
It expresses the conservation of particle pairs separated by a comoving distance $x$ in terms of the two-point correlation function (2PCF) of density fluctuations, $\xi(x, t)$, and the relative velocity of pairs, $v$. Here, $t$ denotes cosmological time, and $a$ stands for the scale factor.  

The result derived from  \refeq{conserv} was employed by \cite{Juszkiewicz_1999} to propose an interpolation-based ansatz for the mean pairwise velocity (MPV), bridging the gap between linear and non-linear regimes, and improving the results proposed by \citep{1991_Hamilton}. The physical reason behind the coefficients of \cite{Juszkiewicz_1999}'s fitting formula was later explained in \cite{2001MNRAS.325.1288S} and extended to treat galaxies in \cite{2001MNRAS.326..463S}.

The BBGKY formalism was used to constrain the cosmic matter density parameter \Om\-\ (as seen in \cite{2000Sci...287..109J}) and for testing cosmological models, including those with non-zero curvature \cite{Juszkiewicz_1999}. However, its performance significantly deteriorates on highly non-linear scales. 

MPVs capture the dynamics of peculiar motions driven by gravitational interactions in an expanding background Universe. This sensitivity has motivated research efforts aimed at using MPVs to identify deviations from GR in large-scale structure formation (as demonstrated in \cite{2014PhRvL.112v1102H,2015MNRAS.451L..45H,2015MNRAS.449.2837G, 2016A&A...595A..40I, 2017MNRAS.467.1386B,2020JCAP...01..055V}). 
The absence of precise modeling for the non-linear regime of pairwise velocities hinders our ability to fully harness the physical information inherent in this observable.

In this paper, we introduce a precise modeling approach for pairwise velocities. Our primary focus lies in enhancing predictions on non-linear scales. We validate our predictions against N-body simulations, and show the versatility and strength of this modelling, by stress-testing it for non-standard cosmologies such as Modified Gravity (MG) scenarios. 

We aim to construct our model using well-established and readily available methodologies rather than introducing new analytical or numerical tools. By demonstrating the effective application of these methods for precise predictions on sub-megaparsec separations, we seek to provide the community with a robust framework for modeling MPVs and using them to test the underlying gravity model. 

This paper is organized as follows: in \refsec{methods} we review the pair conservation equation, describe the models for non-linear scales included in our work and our simulation data. 
In \refsec{results} we present our results for the MPV in \lcdm\-\ and for MG scenarios, which we discuss in \refsec{discussion} and we present our main conclusions in \refsec{conclusions}.


\section{\label{sec:methods} Methods}
We proceed to describe the analytical models, numerical techniques and simulation data we used to build and test our pairwise velocity modelling. 

\subsection{\label{subsec:model}Model}

Our model for computing the mean pairwise velocities, $v_{12}(x,a)$, relies in the consistent numerical solution of the following equation \cite{Juszkiewicz_1999}:

\begin{equation}
    \label{eq:v12full}
    \frac{a}{3[1+\xi(x,a)]}\frac{\partial\bar{\xi}(x,a)}{\partial a} = - \frac{v_{12}(x,a)}{H(a)r}
\end{equation}
where $\bar{\xi}(r,a) = 3x^{-3}\int^x_0\xi(y,a)y^2dy$ is the volume-averaged 2PCF, $H(a)$ is the Hubble function, and  $r=ax$ is the proper separation between pairs.  

Before describing the different approaches for the non-linear clustering, let us briefly examine the limiting values of \refeq{v12full}. In the case of  pairs separated at $x\ll1$\Mpch{} or, equivalently, high values of $\xi \gg1$ we recover the \emph{stable clustering} regime. 
This regime consists of scales where the natural tendency of a pair to collapse, as induced by the gravitational clustering, dominates over the background Hubble expansion, thus a regime where the term $(v_{12}(r)+Hr)$ is negative.
The opposite case, \emph{i.e.} at large $x$ where $\xi(x,t)\ll1$, is known as the \emph{linear-regime} (see for e.g. \citep{1980lssu.book}). 
Here the growing mode of structure formation dominates and we have $\xi(r, t)=D(t)^{2} \xi(r,t=0)=D(t)^{2} \xi_{0}(r)$, with which we get:
\begin{equation}\label{eq:v12lin}
v_{12}(x,a)=-\frac{2}{3}axHf\bar{\bar{\xi}}(x,a),
\end{equation}
where $\bar{\bar{\xi}}(x,a)\equiv \bar{\xi}(x,a)/[1+\xi(x,a)]$ and $f$ is the logarithmic derivative of the growth function, $f\equiv d\ln D/d\ln a$. It is important to recall that the splitting of $\xi(r, t)$ in a time, $D(t)$, and a space dependent function, $\xi_0(r)$, requires that the linear growth function is scale independent, which is generally satisfied in the GR and nDGP cases.

An essential aspect is the transition point where the influence of the Hubble flow on large scales gives way to the stable clustering regime at smaller separations, and then further into  the virialized regime,  where the orbits inside the collapsed haloes are essentially randomized. These transitions are of particular significance, especially when considering cosmologies with different growth-rate histories, such as  MG theories.

\subsection{\label{subsec:MG} Modified gravity}

In our study, as a way of stress-testing our model, we include non-\lcdm{} cosmologies. In particular, we explore MG models as extensions to \lcdm{}. These MG theories introduce alterations to the standard framework, leading to distinct phenomenological consequences that are relevant for our research. Specifically, we focus on two MG theories:
\begin{enumerate}
    \item The normal branch of the Dvali-Gabadadze-Porrati (nDGP) model. This theory seeks to explain the accelerated expansion of the universe by introducing extra dimensions that influence gravity on large scales while preserving standard General Relativity (GR) at small scales. It does so by introducing a scalar field, called the brane-bending mode, which influences the gravitational interaction on large scales.  This modification allows for a departure from standard gravitational laws at cosmic distances.  \cite{ndgp_2000}. 
    \item The Hu-Sawicki form of $f(R)$-gravity \cite{HS_fR_2007}. This widely adopted extension of the Hilbert-Einstein action introduces a functional dependence of the Ricci scalar, $R$, within the Einstein-Hilbert equations, which characterizes the curvature of space-time. The function $f(R)$ allows for variations in the gravitational force as a function of the curvature so that this modification aims to explain cosmic acceleration without invoking dark energy (for a review see for instance \citep{2007PhRvD..75h3504A,2012arXiv1206.1642J}).
    
\end{enumerate}

To deviate from GR on cosmological scales while respecting both high-density and strong-field regime constraints, these MG theories introduce ``screening mechanisms''. These mechanisms are theoretical concepts designed to reconcile the predictions of MG with experimental observations on cosmological and astrophysical scales, effectively concealing modifications to Einstein's field equations in various environments (the regime of high densities or small distances).  
The first family illustrates the Vainshtein screening mechanism \cite{Vainshtein_1972}, implemented in the nDGP gravity model. 
The Vainshtein mechanism relies on the non-linearity of equations of motion for the scalar field, leading to a suppression of its effects at small scales.  
On the other hand, the specific form of $f(R)$ gravity we consider employs the Chameleon screening mechanism \cite{Khoury2003PRD}. In the Chameleon mechanism, a scalar field possesses a variable effective mass dependent on local matter density, allowing it to be screened in high-density environments while remaining active in low-density regions.
For a thorough review on this type of theories we refer the reader to \citep{2012PhR...513....1C}. 
\subsection{\label{subsec:data}Data}

For calibration and testing the pairwise-velocity model, we use data generated from the suite of DM-only MG $N$-body simulations: \elephant{} (Extended LEnsing PHysics using ANalytic ray Tracing), \citep{ALAM2020_ELEPHANT}, as these simulations provide a good test-bed to study the impact of both the Chameleon and Vainshtein screening mechanisms on the large scale non-linear clustering of matter.

This set of $N$-body simulations assumes a \lcdm{} background, and implements on top of this the solution of the scalar field and modified Einstein equations in the MG models described above: normal branch of the DGP theory, or nDGP, and  the Hu-Sawicki form of $f(R)$. In the first case, the specific values of the extra parameter are: $Hr_c = 1,\ 5$ (referred to as  N1 and N5, respectively), and for the second model, the strength of the modification is codified by the present-day value of the derivative of the $f(R)$ function, $|f_{R, 0}|$, taking the values $10^{-5}$, and $10^{-6}$ (referred to as F5 and F6, respectively).  N1 codifies a model that deviates more strongly from GR than N5, in a similar way just as F5 deviates more from GR than F6. 

For each model and redshift, we have five independent realisations. The different MG models have the same fiducial \lcdm{} background. 
For a detailed explanation of the numerical parameters used in this suite of simulations we refer the reader to Sec. 2.2 from \citep{SHED_1}. 

For the direct calculation of the pairwise velocities in the simulation data, we use the positions and velocities of the dark-matter particles and analyse the snapshots at $z = 0,\ 0.3$ and $0.5$. 

\subsection{\label{subsec:pk}Non-linear Power spectrum}
To model the MPV accurately for sub-Mpc separations, we must address the non-linear clustering component in \refeq{v12full}. To achieve this, we employ various proposals for the non-linear power spectrum, denoted as $P_{\text{nl}}(k,z)$. These $P_{\text{nl}}(k,z)$ are subjected to an inverse Fourier transform (IFT) to provide the corresponding non-linear two-point correlation function (2PCF), denoted as $\xi_{\text{nl}}(r,a)$, as required in \refeq{v12full}.
 
\textbf{Non-linear clustering in \lcdm{}: }
In our study, we test and employ the \code{Halofit} solution which is implemented in the cosmological Boltzmann solver \code{CAMB} \cite{Lewis_2000,Howlett_2012}. We fix the cosmological parameters to the values of the  \elephant{} simulations.   The \code{Halofit} formula is an accurate fitting formula for the non-linear matter power spectrum, presented in \cite{halofit_smith_2003} and re-calibrated in \citep{2012ApJ...761..152T}.
We choose to use the \code{Halofit} model for the \lcdm{} case due to its established accuracy, tested to be up to $\sim5\%$ for $k\leq10$\Mpch{}, and $z\leq2$, for a variety of \lcdm\-\ and wCDM cosmologies (see for instance \cite{2012ApJ...761..152T}). 

Additionally, we also compare our results against the solution from the Convolution Lagrangian Perturbation Theory (CLPT) \cite{2013MNRAS.429.1674C}, which is a non-perturbative resummation of Lagrangian perturbation theory and provides as an output, an estimation for \pairvel{}.   

\textbf{Non-linear clustering in Modified Gravity:}
In the context of MG scenarios, which deviate from the standard paradigm, there is less consensus on the optimal approach for modeling the non-linear power spectrum. Various proposals exist, some of which compute it directly using  N-body simulations \citep{PhysRevD.78.123524, Alam_2021}, perturbation theory \citep{PhysRevD.79.123512}, post-Friedman formalism (PPF) \citep{PhysRevD.76.104043}, or via the spherical collapse model \citep{10.1093/mnras/stv2036}. 
Specifically for $f(R)$ gravity, a MG version of the standard \code{Halofit}, known as ``MG-Halofit'', has been introduced \citep{Zhao_2014}. It is essential to note that these approaches are based on certain assumptions and have limitations, which restrict their applicability and generality.
Instead, we primarily utilize the following two tools: the standard \code{Halofit} solution and \code{MGCAMB}\footnote{\href{https://github.com/HAWinther/FofrFittingFunction}{https://github.com/HAWinther/FofrFittingFunction}} \cite{MGCAMB2011JCAP}, a modified version of \code{CAMB} specifically designed to handle both the $f(R)$ and nDGP gravity models. 
Our choice of background cosmology remains consistent between the \lcdm{} and MG cases, aligning with the setup of the \elephant{} simulations. 
Consequently, the primary differences between GR and the MG models manifest in the growth of perturbations. It is important to stress that  the version of \code{Halofit} that we use is agnostic to MG as it was developed and tested for \lcdm{} and wCDM cosmologies \cite{2012ApJ...761..152T}. An in-depth exploration of the limitations to model the non-linear $P(k)$ for MG models can be found in \cite{2023Gupta}. However, and as we will demonstrate in following sections, albeit its lower precision to model non-linear $P(k)$ in MG models, is already enough for getting percent-level predictions for \pairvel{}.

For a quick comparison with our \code{MGCAMB+Halofit}  solutions, we also include a recent approach proposed by \cite{2023Gupta} to compute the non-linear matter power spectrum in the context of MG. 
This work introduces a non-linear matter power spectrum for MG, $P_{\text{nl, MG}}(k, z)$,  in terms of a \lcdm{} non-linear power spectrum,  $P_{\text{nl}, \Lambda \text{CDM}}(k, z)$, and a halo model response function, $\Upsilon(k,z)$, which has been calibrated against the  \elephant{} suite of simulations. 
The resulting power spectrum as function of scale, $k$, and redshift, $z$, $P_{\text{nl, MG}}(k, z)$ can be expressed as: 
\begin{equation}
\label{eq:pkmg}
    P_{\text{nl, MG}}(k, z) = \Upsilon(k,z) \times P_{\text{nl}, \Lambda \text{CDM}}(k,z),
\end{equation}
where $\Upsilon(k, z)$ represents a function that encapsulates the power deviation of \lcdm{} concerning the specified MG model. For $P_{\text{nl},\Lambda \text{CDM}}(k, z)$ in \refeq{pkmg}, we also employed the \code{Halofit} prediction from \code{CAMB}, consistent with the cosmological parameters used in the \elephant{} simulations.
The specific components of the halo model for MG were detailed in \cite{2023Gupta}, where the authors report a 5\% accuracy up to non-linear scales of $k \lesssim 2.5-3$\Mpch{} in their resulting power spectra.

\begin{figure}
\includegraphics[width=\columnwidth]{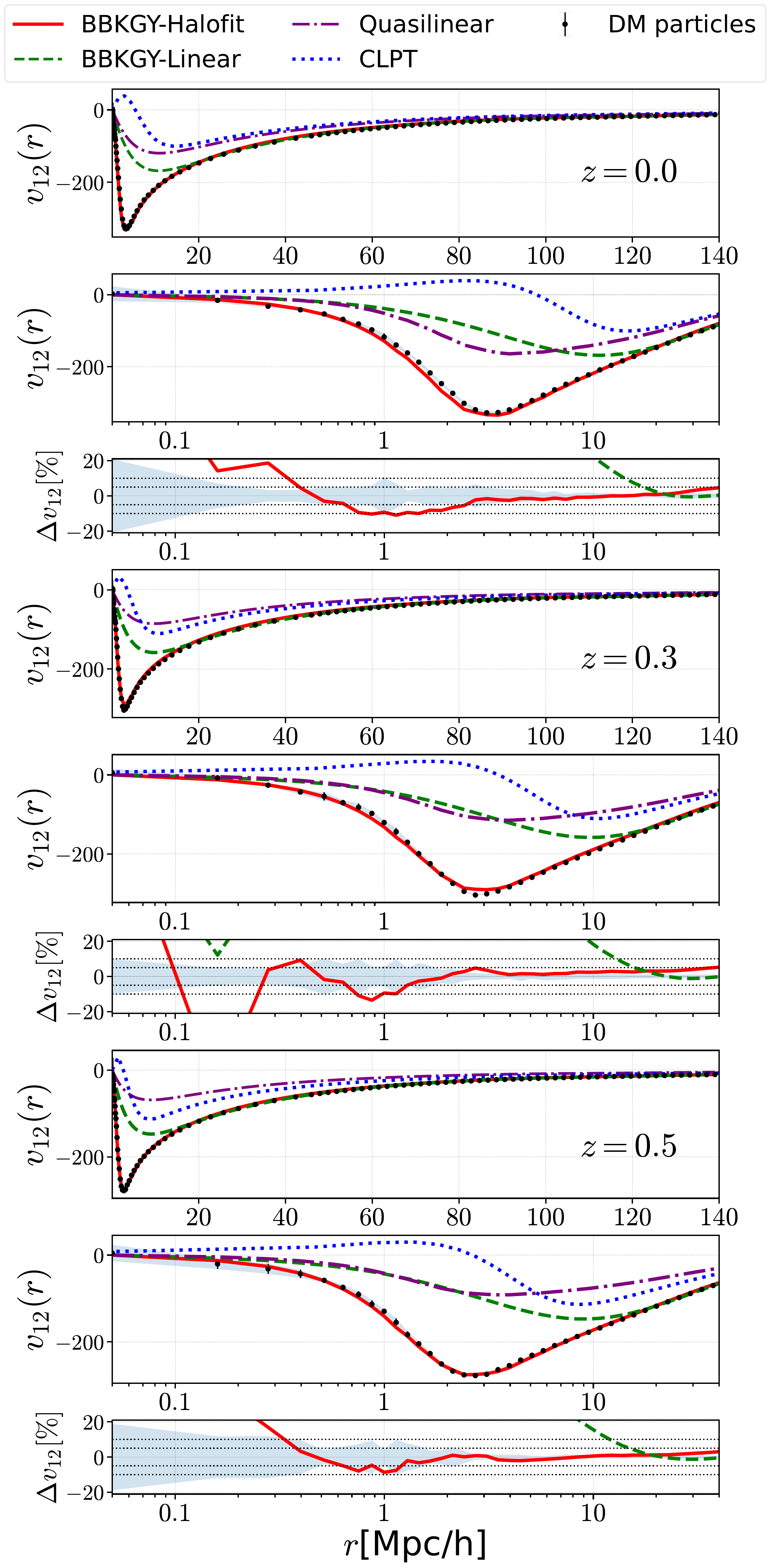}
\caption{\label{fig:v12_GR} Pairwise velocity models for DM tracers, \pairvel{}, in units of $\mathrm{km\,s}^{-1}$ and for the \lcdm{} cosmology. Points correspond to velocities from \elephant{} runs, lines correspond to the different theoretical models: the full solution to \refeq{v12full} using the linear \code{CAMB} matter power spectrum (dashed green), solution obtained via non-linear \code{Halofit} based power spectrum  (solid red), the quasilinear approximation given by eq. \refeq{v12lin} (dot-dashed purple), and the prediction from CLPT (dotted blue).}
\end{figure}



\section{\label{sec:results}Results}

In this section, we present the outcomes of our study, which revolves around the modeling of MPVs in the non-linear regime. To achieve this, we utilize the previously mentioned methodologies for computing the non-linear matter power spectrum and solve for  \pairvel{} using \refeq{v12full}. Our presentation begins with a focus on the \lcdm{} model, followed by an exploration of the applicability of our MPV model in the context of MG scenarios.

\subsection{MPV for the \lcdm{} model.}

Figure \ref{fig:v12_GR} displays the solution for  \refeq{v12full} for \lcdm{} across different panels. Moving from top to bottom, we present the MPV, denoted as \pairvel{} in units of $\mathrm{km\,s}^{-1}$, at varying redshift values. 
The panels labelled with their redshift value: $z=0,\ 0.3\, 0.5$, encompass our findings for separations ranging between $r \in 0.05-140$\Mpch, shown in linear scaling, allowing us to cover both the linear and non-linear regimes simultaneously.
Below these, we zoom in on the region $r \in 0.05-40$\Mpch{}, now in logarithmic x-axis scaling, providing a more detailed view of our solution's behavior in the non-linear regime. Finally, the panels marked with $\Delta v_{12}$, also in logarithmic x-axis scaling, depict the ratio of our numerical solution for \pairvel{} with respect to the simulation data, represented as $\Delta$\vot{}$\equiv (v_{12}-v_{12, \text{sim}})/v_{12, \text{sim}}$.  
The simulation MPVs, $v_{12, \text{sim}}(r)$, are determined directly calculating the projected velocity differences of pairs of dark matter particles. These simulation results are indicated by black dots, with error bars illustrating the variance across five independent realizations of each snapshot.

To obtain numerical solutions for \pairvel{}, we utilize \refeq{v12full} with specific prescriptions. We employ the non-linear output from \code{Halofit}, represented by red solid lines labeled as ``BBKGY-Halofit''. We also include the linear power spectrum from \code{CAMB}, shown as green dashed lines and labeled as ``BBKGY-linear''. 
To facilitate comparison with the results of \cite{Juszkiewicz_1999}, we present a solution to \refeq{v12lin} using the \code{Halofit} power spectrum as input, indicated by purple dot-dashed lines and labeled as the ``quasi-linear'' solution.
Lastly, we include the solution for \pairvel{}  obtained from the numerical implementation of the Convolution Lagrangian Perturbation Theory (\code{CLPT})\footnote{\href{https://github.com/wll745881210/CLPT_GSRSD}{https://github.com/wll745881210/CLPT-GSRSD}}, providing the real-space pairwise in-fall velocity in units of $v/aH(a)f$. These results are represented by the dotted blue lines.

It is worth noting the agreement among all the solutions  for separations $r\geq 80$\Mpch{}. This agreement is expected since these scales fall well within the linear regime, where the dynamics of pairwise velocities are primarily driven by the Hubble expansion. 
It is important to mention that the results from CLPT were originally optimized for galaxy clustering analysis, specifically for extracting the baryonic acoustic oscillations (BAO) feature from the 2PCF of biased tracers. As a result, CLPT provides a robust solution around BAO scales ($r_{BAO} \sim 100$ \Mpch).
However, we observe deviations for separations below $r \sim 80$\Mpch{}. Both the perturbative CLPT and the ``quasi-linear'' solution start to diverge from the simulation data and the other solutions, which becomes especially noticeable in the middle row panel where we focus on separations below $r= 40$\Mpch{}.
On these smaller scales, the CLPT solution follows a trajectory similar to the quasi-linear solution up to $r\sim11$\Mpch{}, below which it deviates the most within the non-linear regime.
The ``BBKGY-linear'' solution performs better than the previous two cases, tracking the evolution of the simulation data within 5$\%$ of accuracy up to separations of $r\sim11$\Mpch{} across all redshift values, but it deviates in the intermediate and non-linear regimes.
Lastly, the fully non-linear solution derived from \refeq{v12full} with $P_{\text{non-lin}}(k)$ from \code{Halofit} closely mirrors the 
full non-linear dynamics of dark matter particles in the simulation, achieving an accuracy of approximately $\sim 10\%$ up to separation of $r \sim 1.1$\Mpch, for $z=0$ and, $r \sim 1$\Mpch, for $z= 0.3, 0.5$.
Our solution exhibits a consistent trend with redshift  when compared to the simulation data, showing a gradual decrease in the \pairvel{} value at its minimum during earlier snapshots. We delve deeper into the analysis of this characteristic within the framework of various MG models.

For the remainder of the paper, unless explicitly specified otherwise, it is presumed that the solutions presented pertain to the ``BBKGY-Halofit'' method.

\begin{figure*}
\includegraphics[width=0.8\textwidth]{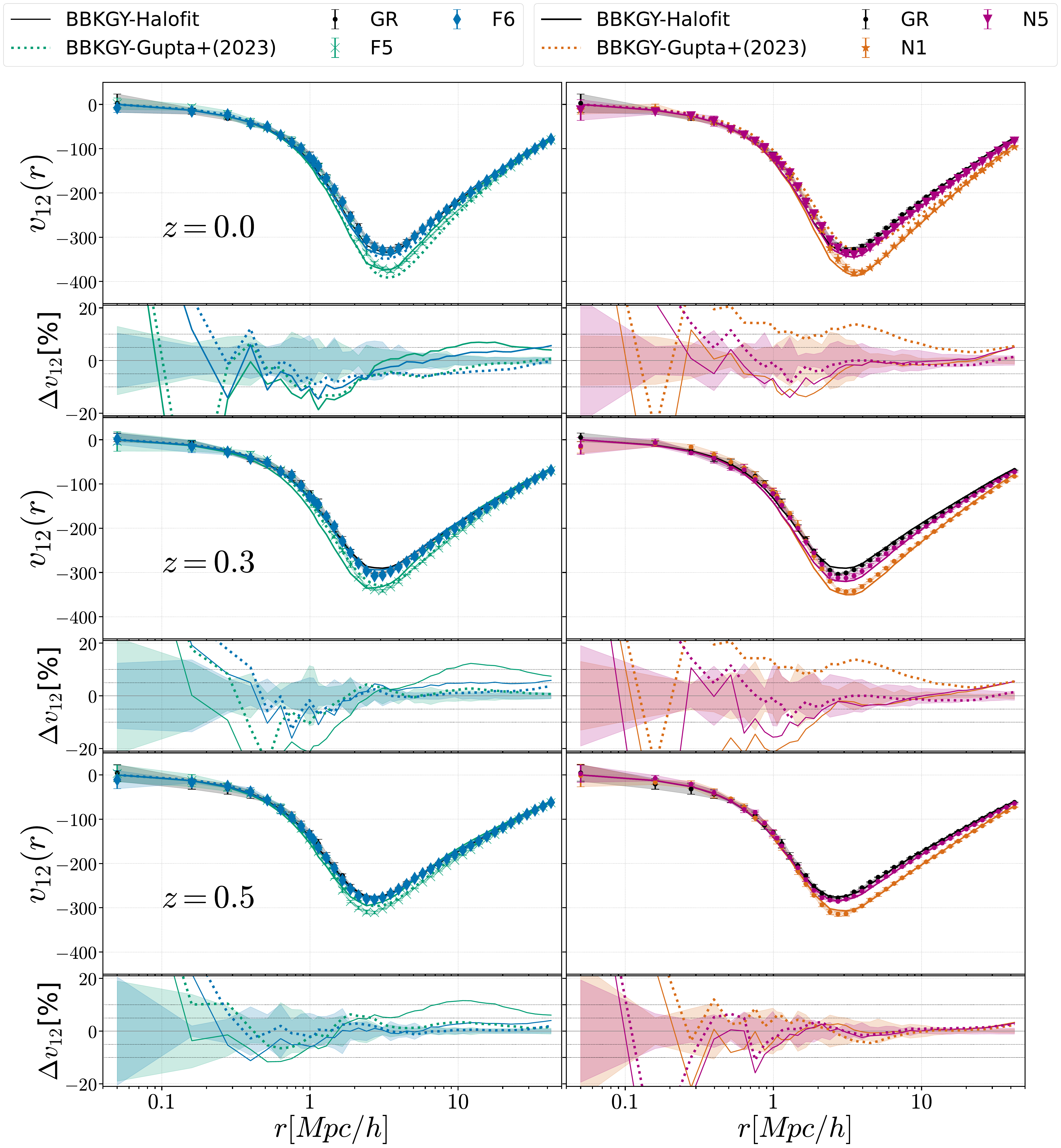}
\caption{\label{fig:v12_MG} Pairwise velocities, \pairvel{} in units of $\mathrm{km\,s}^{-1}$, and models in MG theories. Points correspond to DM particles from the \elephant{} runs, solid lines to the result from BBKGY-\code{Halofit} prediction, and dotted lines to the solution using \refeq{pkmg}, labeled as ``BBKGY-Gupta(2023)''. The results for the GR case are represented in black, the F5 model with turquoise, F6 in blue, N1 in orange and N5 in magenta. }
\end{figure*}

\begin{figure}
    \centering
    \includegraphics[width=1\linewidth]{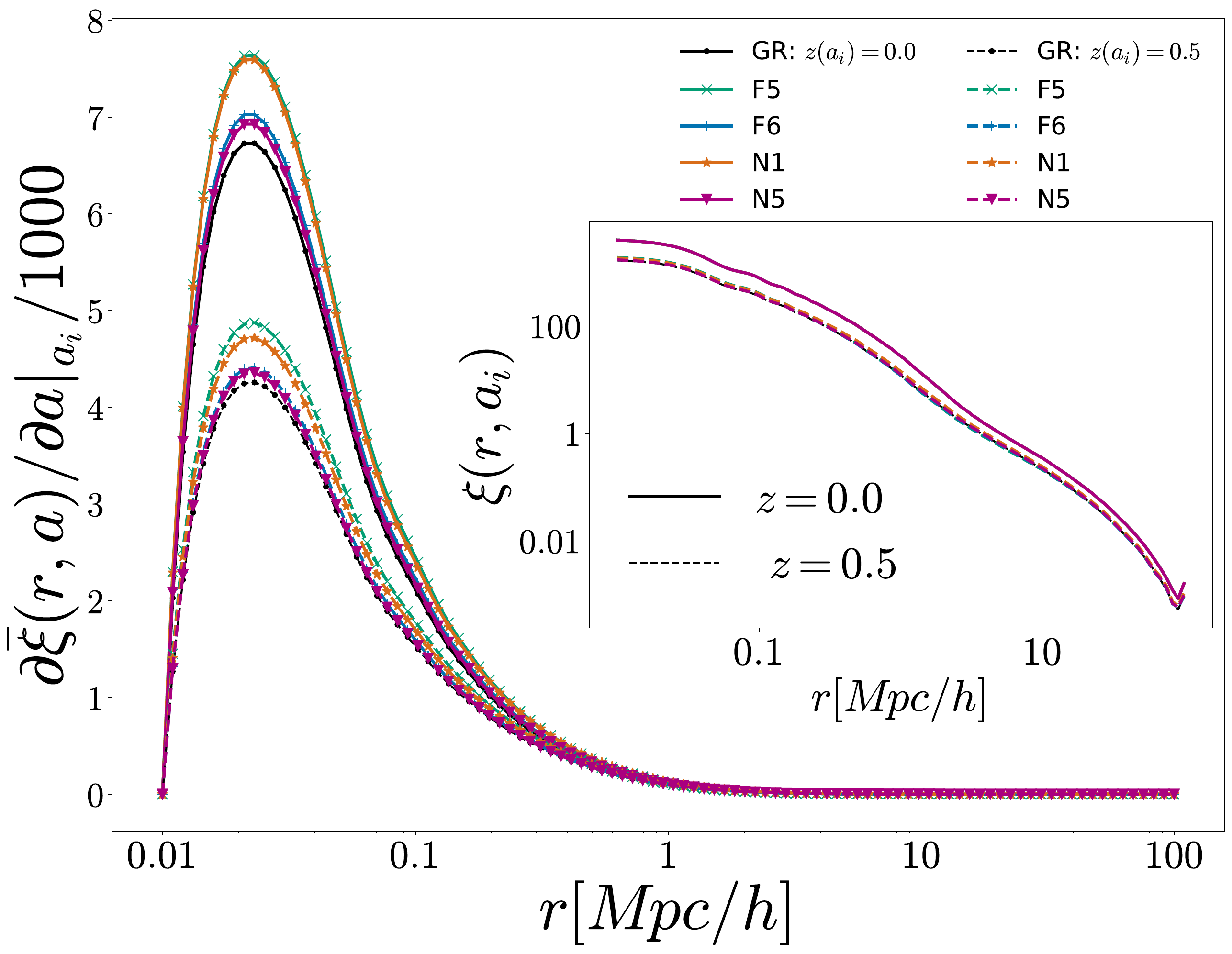}
    \caption{Source (left-hand side) term of \refeq{v12full} with a focus on the  time derivative of the volume-averaged correlation function part, for the different gravity models (GR, F5, F6, N1, and N5), and for $z=0$ (solid lines), $z=0.5$ (dashed lines). As an inset plot we show the factor multiplying $\partial\bar{\xi}/\partial a$, for the same models, $z$ values, and $r$ range. }
    \label{fig:source}
\end{figure}

\subsection{MPV in MG theories.}

In our treatment of  MG theories, the modifications to GR are contained in the linear power spectra from \code{MGCAMB}, which properly takes into account the modified gravitational interactions for a Hu-Sawicky $f(R)$-gravity, and the nDGP model. The non-linear part of the solution, however, is generated from the standard \lcdm{}-\code{Halofit} fitting formulae.  This setting will be referred to as ``BBGKY-\code{Halofit}'' in our presentation. 

The  resulting $v_{12}$ models are visualized in \reffig{v12_MG} in units of $\mathrm{km\,s}^{-1}$, which, as done previously for the \lcdm{} case,  consists of three rows corresponding to three snapshots from the simulations: from top to bottom, $z=0,\ 0.3,\ 0.5$. 
The rows labeled with the corresponding snapshot value showcase the trend of \pairvel{} in these MG models over the range of separations $r\in[0.05, 40]$\Mpch{}. 
Specifically, the six panels on the left illustrate the results for $f(R)$ variants (F5 and F6), while the six panels on the right depict the results for the nDGP variants (N1 and N5). 
The data points with associated errors represent the simulation outputs obtained directly calculating the projected velocity differences of  pairs of dark matter particles. 
To facilitate comparison, we also present the simulation results for GR in all the figures, denoted by the solid points, as well as the corresponding \code{Halofit} solutions to \refeq{v12full}, in black lines.
 
In each case, the bottom sub-panel displays the ratio of the solution for \pairvel{} with respect to the corresponding MG model simulation, expressed as $\Delta$\vot{}$\equiv (v_{12}-v_{12, \text{sim}})/v_{12, \text{sim}}$. 
The shaded regions in these subplots represent the corresponding relative errors associated with the simulation data.

Starting with the plots  which illustrate the full evolution of \pairvel{}, we observe that the simulation data for the F6 (blue points) and N5 (magenta points) models closely track the evolution seen in the GR case (black points). This trend persists as we move from top to bottom in \reffig{v12_MG}, indicating that our model captures the temporal evolution even in these non-\lcdm{} scenarios.

Regarding our model for \pairvel{} (solid lines), we see how our solution from the ``BBGKY-Halofit'' model captures the trends of the simulation data in all cases and for all snapshots. For the case of Hu-Sawicki $f(R)$ gravity, we find that our model follows more closely the simulation data in the weaker version of the theory (F6, indicated by green lines and green points). For the F5 variant, we notice that our solutions deviate from the simulation data at the larger separations $r >10, 5, 1 $ \Mpch{}, for the snapshots $z=0.5, 0.3, 0$, respectively. 
To better appreciate how the dynamics of pairwise velocities is captured by our model, we refer to the ratios $\Delta v_{12}$ shown in the narrow subplots at each panel. 
Starting from the left side, or the $f(R)$ model, we notice a constant offset of the ``BBKGY-Halofit" solution at large scales ($r \approx 40$ \Mpch). This offset reaches $\Delta v_{12}\sim 10 \%$ for the F6 case, and it is more prominent ($\Delta v_{12}\sim 20\%$) for the stronger variant of the model, F5. In contrast, the solutions obtained with \refeq{pkmg} or ``BBKGY-Gupta(2023)",  recover the expected values at these large separations. 

Generally speaking, our model with the ``BBKGY-Halofit" solution accurately captures the dynamics from the $f(R)$ simulations with better than 10$\%$ accuracy for separations of a few \Mpch{} (between $2-10$\Mpch{}). 
However, as we approach the fully non-linear regime or sub-megaparsec separations, there are noticeable deviations from the data, particularly for F5 at $z = 0.3$. 
The results from using the ``BBKGY-Gupta(2023)" prescription perform better in all the cases mentioned above.

We notice how both solutions perform well for the $z = 0.5$ snapshot, achieving an accuracy of of $10\%$ or better for scales down to $r\sim0.3\ (0.15)$\Mpch{}, for F5 and F6, respectively. 
 
Turning to our nDGP solutions (right column of \reffig{v12_MG}), we observe a better agreement between our model and the simulation data, for the specific case of the ``BBKGY-Halofit" solution. In each sub-panel presenting the ratio $\Delta v_{12}$, we see that our prediction initially shows an offset of $\sim 10\%$ at $40$\Mpch{} separations but then converges to the result from direct calculation in the simulation data for separations between a few and $10$ \Mpch{}. At  $z = 0$ our model captures the non-linear evolution of \pairvel{} with 10$\%$ accuracy or better for pairs separated between $3-40$ \Mpch{}. This level of accuracy is similarly observed at $z=0.3$, and it improves further at $z=0.5$, where we achieve an agreement better than 10$\%$  for scales $r\geq1$\Mpch{}.
For our model using the ``BBKGY-Gupta(2023)" prescription (shown in dotted lines) we notice that these solutions under-perform in comparison to the ``BBKGY-Halofit" solution, particularly for $z=0.3$, at which the pairwise velocities offset by 10$\%$ or more the direct calculation from our simulations, for separations $r\sim5$\Mpch{}.

To understand better where these differences among gravity models come from, in \reffig{source} we illustrate the various components of the source term (left-hand side) of \refeq{v12full}. 
The clustering itself, denoted by $\xi(r,a)$, exhibits minimal variation among gravity models (depicted in the inset figure for redshift values, $z=0$, and $0.5$, for variants F5, F6, N1, N5, and the GR case). 
However, its time evolution, represented by $\partial\bar\xi/\partial a$, reveals notable differences across models.

Specifically, the more pronounced modifications of GR, N1, and F5 exhibit a maximum deviation of approximately $13\%$ from GR, at $z=0$ (shown in solid lines), in contrast to the milder variants (N5 and F6), which deviate around $3-4\%$ from the GR prediction at their maximum. 

Notably, while at $z=0$, the values for $\partial\bar\xi/\partial a$ for models F6 and N1 coincide, these models display different values at $z=0.5$. In this case, their relative deviations from GR are approximately  $11\%$ and $14\%$, respectively, while the N5 and F6 variants show deviations of around $2-4\%$ compared to the GR prediction.

\subsection{\label{subsec:mgsignatures} Signatures of MG in the MPVs}

Comparing pairwise dynamics across different gravity models reveals interesting characteristics.
As we can see in \reffig{v12_MG}, the minimum value of \pairvel{} undergoes variations based on the gravity model. Specifically, for the stronger deviations from GR, \pairvel{} reaches a more pronounced minimum compared to scenarios involving weaker modifications of gravity or adhering to the GR framework.

\begin{figure}
  \centering
\includegraphics[width=0.5\textwidth]{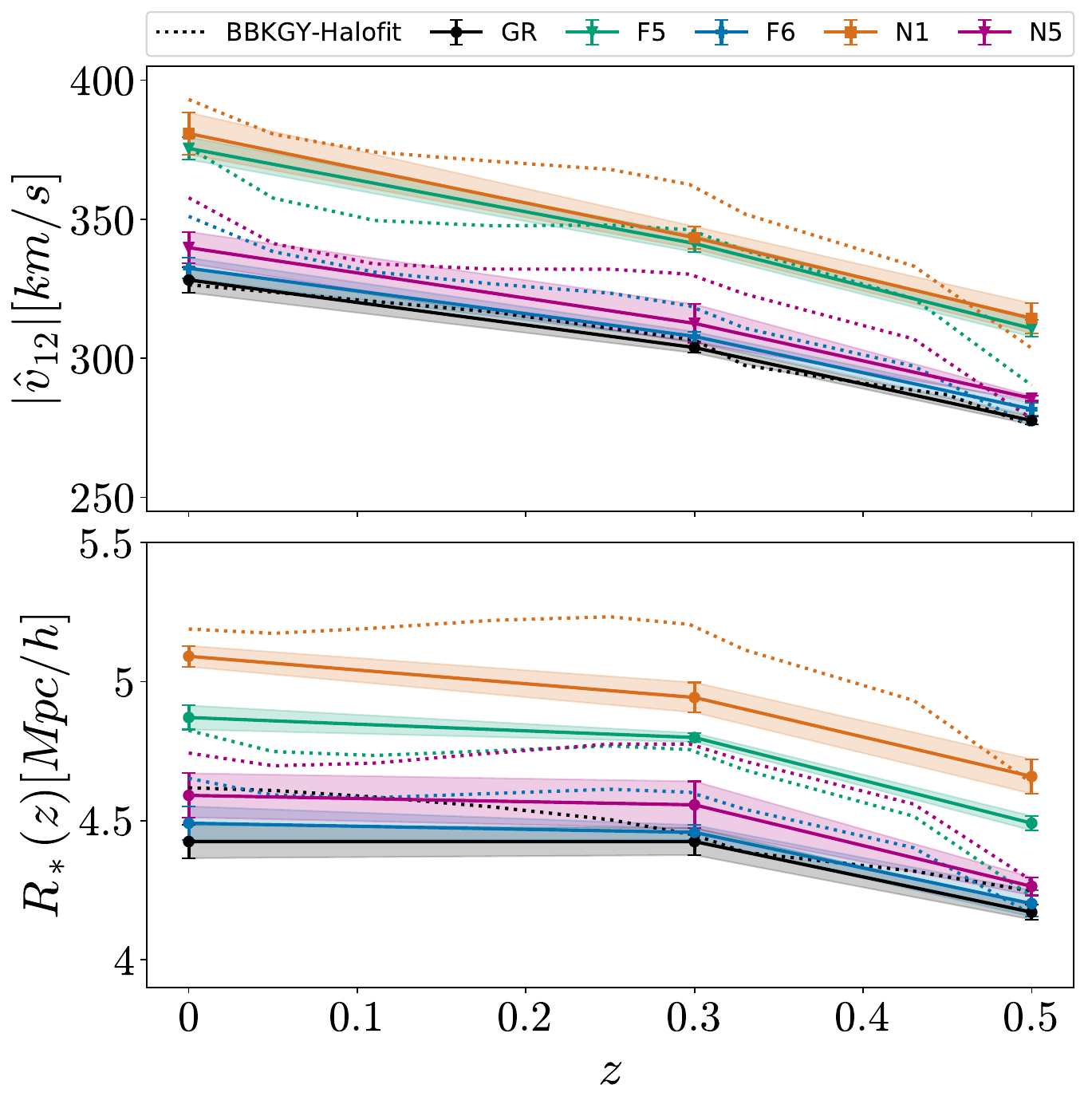}
    \caption{Prediction of the maximum value of the infall pairwise velocity, $|\hat{v}_{12}(z)|$, and the stable-clustering crossing scale, $R_*$, as function of redshift. The expectation of a decreasing function for higher values of $z$ is present for all models; however, we notice differences compared to GR. }
      \label{fig:rstarpeakv12}
\end{figure}

To quantify this effect, we present the maximum value of $|$\pairvel{}$|$, denoted as $|\hat{v}_{12}|$, as a function of redshift, $z$, in the top row of \reffig{rstarpeakv12}. As anticipated, during earlier cosmic times, relative velocities attain smaller magnitudes in comparison to the present epoch. Across all cases displayed in the top row of \reffig{rstarpeakv12}, there is a consistent trend toward smaller $|\hat{v}_{12}|$ (indicated by shallower curves for \pairvel{} in earlier snapshots, as seen in \reffigs{v12_GR}{v12_MG}). The noteworthy observation lies in how this value varies across distinct gravity models.  

First we focus on the values from the simulation data. These are represented by the error-bar points at each snapshot value: $|\hat{v}_{12}| (z=0,\ 0.3,\ 0.5)$. The errors express the standard deviation relative to the mean calculated from all realizations, also shown in  the respective shaded regions. 
The distinctions between the values of $|\hat{v}_{12}|$ for GR and MG models become quite apparent. 
While the value of $|\hat{v}_{12}(z)|$ for the weaker modifications of GR  (the variants F6 and N5 denoted by blue crosses and purple diamonds, respectively) closely tracks the trend of GR (indicated by black circles), we see that for the more pronounced MG variants (F5 and N1, represented by green triangles and orange squares, respectively) $|\hat{v}_{12}(z)|$ exhibits significant deviations from the GR values. 
In particular, at $z=0$, the relative difference, defined as $\Delta|\hat{v}_{12}| \equiv (|\hat{v}_{12,\text{MG}}|/|\hat{v}_{12,\text{GR}}|-1)$, reaches $16\%$ (or an increment of $52 \mathrm{km\,s}^{-1}$), for the N1 case. Similarly, the F5 variant has $\Delta|\hat{v}_{12}| \sim 14\%$ with respect to the GR value (an increment of $\sim 47 \mathrm{km/s}$).   
For the models N5 and F6, this deviations are negligible with respect of the uncertainties ($\Delta\hat{v}_{12} \approx 3.6, 1.2 \%$, respectively).
For earlier snapshots, the deviation between N1 or F5 and GR is of the same order of magnitude: $\Delta|\hat{v}_{12}|  \approx 12-13\%$. Importantly, these deviations between gravity models are larger than the uncertainties in $\Delta|\hat{v}_{12}|$ from the simulation data. 
Conversely, for the weaker modifications of GR, as expressed in the F6 and N5 variants, the relative deviations are contained within the data uncertainties.  
We add the solutions from our model, shown as dotted lines. 

\begin{figure}
    \centering
    \includegraphics[width=0.9\linewidth]{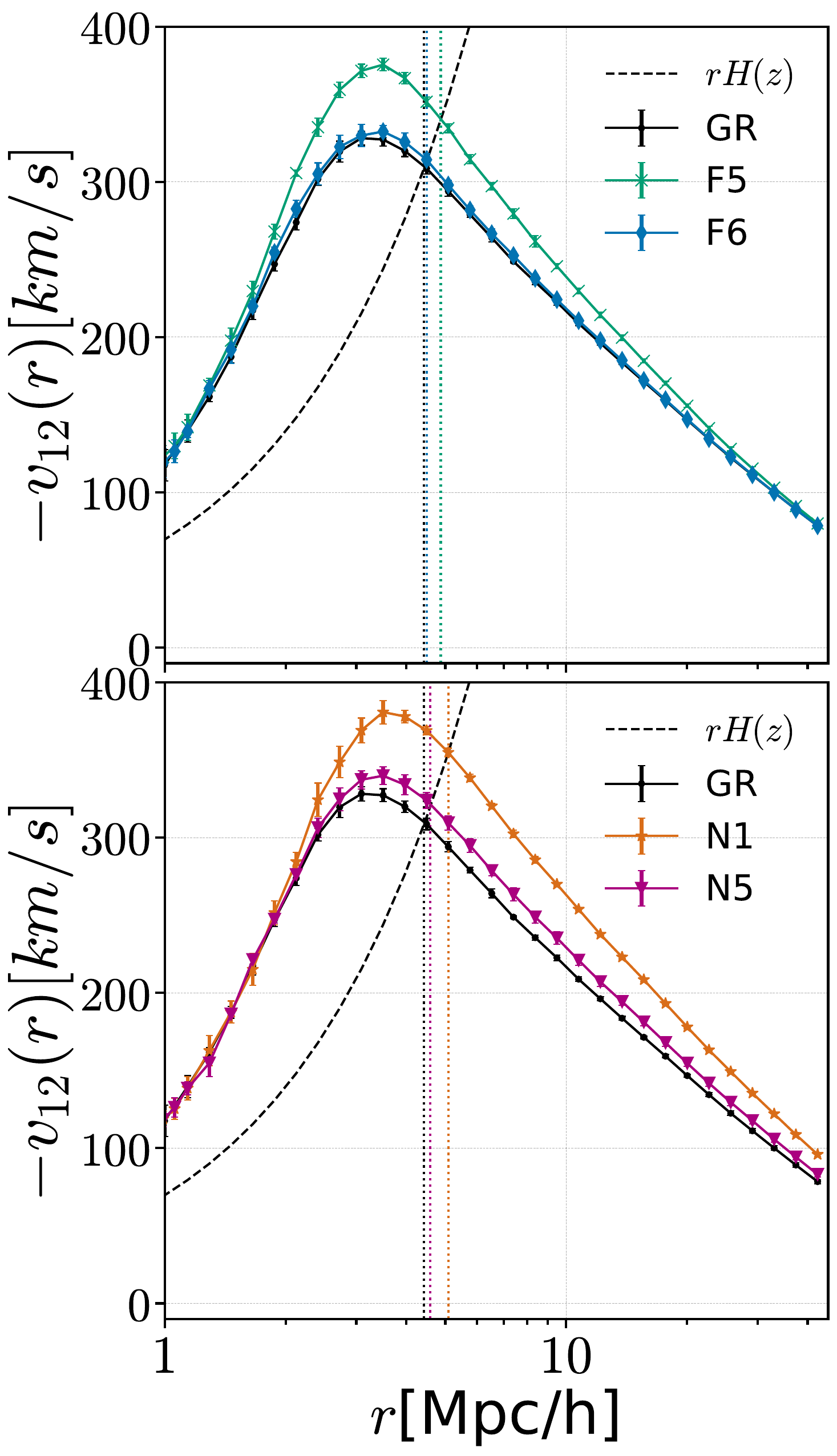}
    \caption{Stable-clustering crossing scale, $R_*$, at $z=0$ for the different models used in this work. In both panels we show the term $-v_{12}(r)$ for each model, and the Hubble flow $rH(z)$, which is identical in all models. 
    The data points represent the values extracted from the simulation data with error-bars displaying the variance from five independent box realizations. 
    In the top panel we show the \pairvel{} for the F5 and F6 variants of $f(R)$-gravity, and the bottom panel shows the N1 and N5 variants of the nDGP model. }
    \label{fig:rstarhubble}  
\end{figure}

For the GR case, and the N1 and F5 variants, our prediction for $|\hat{v}_{12}|$ agrees within the data uncertainties, except for N1 at $z=0.3$, while our prediction for $|\hat{v}_{12}|$ in the case F6 and N5 presents an offset with respect to the simulation data. However, in all cases we recover the two main trends shown in our data: a decreasing value of $|\hat{v}_{12}|$ for earlier snapshots, and more importantly, the relative increment with respect to GR in the different MG scenarios. 
In particular, we recover the maximum deviation for the N1 variant, followed by the F5 case, then the N5 variant, and lastly, the F6 model.

Another interesting aspect to consider is the scale at which each model transitions into the stable clustering regime. We denote this specific scale as $R_*$ and defined to be such that $aR_*H(z)-v_{12}(R_*)=0$. To illustrate this point, in \reffig{rstarhubble}, we explicitly show the values for $-v_{12}(r, z=0)$ for all the models under study, together with the Hubble velocity, $rH_0$ (dashed line). In vertical dotted lines we indicate the value of $r$ at which both lines cross each other, in other words, the value for $r=R_*$ for each model. 

As previously explained, \pairvel{} is dominated by the Hubble flow at large separations, gradually transitioning into the stable clustering regime as pairs of objects draw closer (\citep{1991_Hamilton,2001MNRAS.325.1288S}). However, the scale $R_*$ at which this transition takes place varies depending on the underlying gravity theory. The outcomes of this analysis are presented in the bottom panel of \reffig{rstarpeakv12}.

In the bottom panel of \reffig{rstarpeakv12} we show the stable-clustering crossing scale $R_*(z)$ for each gravity model. The values from our simulation data are shown in solid lines with error bars and the corresponding model prediction, in dotted lines. 
An interesting result is how this scale increases for the MG models that deviate more prominently from the GR case.   We recover a similar behaviour as the variation of $|\hat{v}_{12}|$ with the gravity model, namely, that the strongest deviation in the value of $R_*$ is found for the N1 variant, followed by the F5 case, the N5 and lastly, the F6 case. Our model predictions lie outside the data uncertainties. Nevertheless, we recover the data trends for each one of the MG cases.

\section{\label{sec:discussion}Discussion}

In this paper we have presented the modeling for the evolution of pairwise velocities, for  the linear, quasilinear, and fully non-linear regime. Our model relies on the proper solution of the clustering in said regimes, and at its core, it sits on the pair-conservation equation as derived from the BBGKY hierarchy. 
The elegance of this approach allowed us to test its validity beyond the standard framework of GR and \lcdm{}. 

In  \reffig{v12_GR} we have shown the solution from \refeq{v12full} and compared it against known approximations, and also against the direct calculation of the projected relative velocity differences of pairs in our N-body simulations. 

In particular we see that the fully consistent solution of \refeq{v12full}, using the \code{Halofit} template for the the clustering (``BBKGY-Halofit''), provides excellent agreement with the \lcdm{} simulation data throughout all the range ($r\in[0.05-140]$\Mpch{}), and importantly so, in the sub-megaparsec regime.  

It is essential to acknowledge the limitations of our direct estimation from simulation data, with the smallest resolved scales being at $r=0.05$\Mpch{}. A recent study \citep{2023MNRAS.525.1039M} delved into the accuracy of pairwise velocities in N-body simulations, albeit in a scale-invariant Einstein-de-Sitter cosmology ($\Omega_{tot}=\Omega_m=1$). Despite this limitation, their findings revealed that the direct calculation of projected velocities between particle pairs in the simulation and the estimation using the 2PCF in the conservation equation are equivalent.  Importantly, their simulations reached smaller scales than those covered in our work, which encompasses the range of scales presented in our analysis. 

While perturbative solutions, such as CLPT, were proposed to provide accurate solutions in the BAO regime, they understandably do not match the evolution of the \pairvel{} below several tens of \Mpch{}. 
The ansatz proposed by the authors of \citep{Juszkiewicz_1999} (labeled in our solutions as ``Quasilinear'')  shows a better performance in comparison to the perturbative treatment of CLPT. While the behaviour of \pairvel{} for CLPT shows an nonphysical turn towards positive values of \pairvel{} in the range of $r\leq10$\Mpch{}, the ``Quasilinear'' approximation correctly shows the tendency to a decreasing value in the \pairvel{} in such regime. Importantly too, we can notice how the ``Quasilinear'' converges to the solution of the full BBKGY hierarchy using as an input a linear model for the  2PCF (a solution we named ``BBKGY-linear"). This result shows the power of consistently solving \refeq{v12full}, and its ability to provide better results than with other approximations. 

In the bottom panel of \reffig{v12_GR} we show that our numerical solution is within $10\%$ accuracy  for  the three snapshots, all the way down to $r\approx1$\Mpch. In the particular case of $z=0.5$, however, our numerical solution is consistent with numerical errors (which are below the $10\%$ lines) even at separations of $r\sim 300-400$\kpchInv{}. For the intermediate snapshot, $z=0.3$, this is achieved up to $r\sim 1$\Mpch, and for  $z=0$, up to $r\sim2$\Mpch{}. 

The underlying  assumptions for the validity of \refeq{v12full} are tied to the conservation of pairs of dark matter particles in an expanding Universe (see chapter IV of \citep{1980lssu.book} and the discussion in \cite{2001MNRAS.326..463S}). We check whether the solutions can be applied to  cosmological models beyond \lcdm{}. In particular, we have investigated the case when we relax the assumption of General Relativity as the underlying law of gravity at all scales. 

In \reffig{v12_MG} we can see the remarkable agreement between the solution of \refeq{v12full} and the simulation data obtained for the MG scenarios described in \refsec{methods}.  In this case we present the results of fully solving the BBKGY hierarchy \refeq{v12full} with the \code{Halofit} prescription to obtain the non-linear power spectrum, $P_{\text{nl}}(k, z)$, from the linear power spectrum in the MG scenarios under consideration, $P_{\text{lin, MG}}(k)$.  
While, as mentioned previously, the \code{Halofit} ingredients were calibrated to match the non-linear clustering in \lcdm{}, our $P_{\text{lin, MG}}(k)$ properly takes into account the modified gravitational interactions for a Hu-Sawicky $f(R)$-gravity, and the nDGP model. 

In the first MG scenario, the Hu-Sawicki $f(R)$ theory, we found a constant shift between our solution and the simulation data, for pairs separated a few tens of \Mpch{}.  
While one plausible explanation for this deviation might stem from the absence of a well-defined large-scale limit for pairwise velocities, as discussed in \cite{2004PhRvD..70h3007S}, an alternative hypothesis implicates the potential influence of the absent effective screening mechanism in this intermediate regime (at approximately 
($r=30$\Mpch{}). The observed constant shift prompts speculation about the nuanced interplay between the characteristics of the 
$f(R)$ theory and the larger-scale dynamics, necessitating further investigation to discern the precise origins of this discrepancy.

Note however that this offset is not present when we use the solution from \citep{2023Gupta} for the $f(R)$ model. Instead, the regime of applicability of our model is extended using the ``BBGKY-Gupta'' solution in our prediction for \pairvel{}. We recall that the response function in \refeq{pkmg} was calibrated against the \elephant{} suite of simulations, to provide a proper modeling for the non-linear clustering in the MG scenarios under consideration. 

On the other hand, for the nDGP case (bottom panel of the second row of \reffig{v12_MG}) we show that our solution properly models the dynamics for pairs separated $r\gtrapprox2$\Mpch{} for the snapshots $z=0,\ 0.3$, and for pairs separated at  $r\gtrapprox1$\Mpch, for $z=0.5$. We compare our model against the numerical results obtained from the simulation data, and shown as shaded regions around the value $\Delta$\pairvel{}$\equiv \left[ v_{12}-v_{12,\text{sim}}\right]/v_{12,\text{sim}}$. This level of agreement is achieved with our ``BBKGY-Halofit" prediction. In the case of the ``BBKGY-Gupta(2023)" solution the results deviate beyond the propagated error range of the simulation data across all values of $r$, in the specific case of N1, at the snapshots $z=0, 0.3$, but we obtained a better match for the value $z=0.5$. The solution for the N5 variant from the ``BBKGY-Gupta(2023)" was found more consistent with the uncertainties from the simulation particles in the three $z$ values. We speculate that this can be an effect of the offset in their response functions (figure 2 of \citep{2023Gupta}) for the specific case of N1.

In all the cases that we discussed, both in GR and MG cosmologies, we find better agreement between our solutions and the simulation data for the higher redshift snapshot ($z=0.5$), in comparison to the subsequent values, $z=0.3, 0$. 
This can be attributed to the fact that the more recent snapshots represent highly non-linear stages of the evolution of the Universe, therefore, the scale at which our prediction is accurate decreases with the redshift. 

From examining the left hand side term of \refeq{v12full}, we conclude that it is the time evolution of $\bar{\xi}$ what determines the dependence of \pairvel{} on the gravity model.  

A perhaps more interesting result can be seen in the results shown in \reffig{rstarpeakv12}, where we study the differences in the dynamics of \pairvel{} between the various gravity prescriptions.  
We focus on two distinct features, the value of \pairvel{} at its minimum as a function of redshift, $\hat{v}_{12}(z)$, and the scale at which the pairs enter the stable clustering regime, $R_*(z)$, instead of being dominated by the Hubble flow. 
This scale is defined by the ratio of the streaming velocity to the Hubble expansion. It is important to remember that the modifications of the underlying gravity model are taken into consideration only at perturbation level in the \elephant{} suite of simulations, keeping the same background as in GR for all the models under scrutiny. In other words, the effect of changes in the expansion rates that such models have (see for instance \cite{PhysRevD.89.084010,2018PhRvD..98h3530J,2022PDU....3701069J} where the specific case of $f(R)$-gravity theories in the cosmological context is discussed) are not considered in our simulations.
Therefore, the differences found in the value  of $R_*(z)$  express changes arising only from gravitational clustering in these MG scenarios, which helps to pinpointing changes attributed to the modified force law rather than by the modified expansion dynamics. The same claim is valid for the maximum value of  $|v_{12}(z)|$, denoted as $|\hat{v}_{12}(z)|$, which encodes the enhanced  infall velocities of pairs of galaxies in this alternative gravity models, with respect to GR. 

Interestingly, we see a connection between the effect of the strength of the modification of gravity and the changes in the values of $|\hat{v}_{12}(z)|$ for a given $z$: $\Delta |\hat{v}_{12}|\equiv|\hat{v}_{12,\text{MG}}|/|\hat{v}_{12,\text{GR}}|$. 
The  stronger modification of GR in the nDGP model, the case N1, displays the larger difference $\Delta |\hat{v}_{12}|$, followed closely by the F5 case of the $f(R)$ model. 

These findings point to a distinct and recognizable signature from modifications of gravity on cosmological scales, which can potentially help detect the clear effect predicted by the implementation of these models, supporting the results presented in the letter \citep{2014PhRvL.112v1102H}, where different statistics for the pairwise velocities were analyzed. In that work, the authors focus on the amplitude of $\sigma_{12}$, the  line-of-sight centered pairwise dispersion, also derived under the BBKGY formalism.  A direct comparison to their results is not possible, as their analysis was based on the HOD (Halo Occupation Distribution) mock galaxies, while we have kept our analysis on the dark matter particles, for which we are guaranteed that the conservation of pairs is fulfilled throughout cosmic history in the different snapshots we analysed. 
However, some indirect comparison can be made as a part of the signal in $\sigma_{12}$ that originates, in fact, from $\Delta$\pairvel{}. Their results showed an increase in the amplitude of $\sigma_{12}$, for F5 with respect to GR, of approximately $25\%$ at separations of $r=1$\Mpch{} and $r=5$\Mpch{}, the two cases probed in their analysis. 
This reinforces the primary finding that pairwise velocities serve as a potent tool for assessing the validity of General Relativity (GR) on cosmological scales. 


\section{\label{sec:conclusions}Conclusions}
In summary, this study has outlined a robust methodology for accurately computing pairwise velocities across a wide range of regimes, encompassing linear, mildly non-linear, and fully non-linear stages. We have demonstrated that by adequately considering the clustering aspects and their temporal evolution, our approach can effectively predict the mean pairwise velocities  (MPVs) within the range covered by our clustering model. Our analysis relies on the fundamental equation derived from the BBGKY hierarchy and employs non-linear power spectrum models. Through Fourier transformation, we obtain the non-linear two-point correlation function $\xi_{\text{non-lin}}(r,a)$, which serves as the input to our core equation.

It is important to emphasize that this approach does not necessitate simplifications or approximations to generate reliable predictions for \pairvel{}.

Furthermore, we have shown that this model is applicable to gravity models other than GR.
Specifically, we have established that by appropriately accounting for non-linear clustering in these alternative gravity models, our equation can seamlessly provide predictions for the infall velocities of pairs in these modified gravity (MG) scenarios. This highlights the versatility and robustness of our methodology, making it suitable for a broader range of cosmological investigations.

Even more, the different MG scenarios have a particular imprint in the dynamics of in-fall velocities. We have shown that the physical scale at which the dynamics between pairs is dominated by their gravitational attraction, as opposed to being dominated by the Hubble flow (crossing scale in the stable-clustering regime, $R_*$), is distinctively affected by the MG families of theories we analysed. As screening mechanisms are a relatively generic prediction of viable MG theories,  detecting deviations in the pairwise dynamics as we have described them would be a signature of physics beyond GR. 

Furthermore, the pairwise velocity dispersion (PVD), $\sigma_{12}$, appears in the first moment of the second BBKGY equation \citep{1971phco.book.Peebles,2014PhRvL.112v1102H}.
The analysis of PVD is a vital component of redshift space distortions (RSD) models.  To harness the full potential of ongoing and future data collection cosmological surveys like DESI \citep{Desi}, Euclid \citep{laureijs2011euclid}, and 4MOST surveys (CRS \cite{4most}, and 4HS \cite{4HS}), we must push the boundaries of our current RSD models. An extension of our model could result in physically-motivated model for RSD that could be applicable to a variety of gravity theories. 

However, as we delve into smaller separations, the interplay between baryonic physics and dark matter distribution becomes increasingly significant. Specifically, the effect of baryons on RSDs, cosmic density and velocity fields has been thoroughly investigated in \cite{2016MNRAS.461L..11H} by employing comprehensive hydrodynamical simulations. 
Their findings reveal that the impact of baryonic matter on halo and galaxy velocities becomes notable only at very small separations, typically in the order of kiloparsecs. In fact, for separation scales smaller than $r\sim0.6$\Mpch{}, the deviation in the total matter power spectrum is anticipated to be less than $20\%$. Furthermore, over a broad range of distances spanning from $r\sim1.6$ to $62$ \Mpch{}, the overall amplitude of the power spectrum deviates in less than $1\%$ when compared to DM-only simulations. Therefore, the impact of baryons on the velocities of the DM samples employed in our work remains negligible over the broad range of scales under consideration. 

However, for a realistic detection of such signatures we need to consider a number of systematic effects.  The most immediate one is that we have modelled the signal directly on the pairs of dark matter particles, which can be guaranteed to be a closed system therefore ensuring the conservation of pairs. Conversely, halos and galaxies evolve through mergers and accretion, leading to a number density that is a stochastic function of the underlying dark matter distribution—a relationship known as bias. While on large scales, we can expect a linear bias relationship, we aim to describe the dynamics of streaming motions on a wide range of scales. Therefore, a non-linear bias model is crucial to properly account for the dynamics of pairs at sub-megaparsec scales. Although the studies by \cite{2001MNRAS.325.1288S,2001MNRAS.326..463S} have laid the groundwork in this area, the scenario involving theories with scale-dependent growth, and therefore, a scale-dependent bias, needs further exploration.  These are immediate steps we plan to tackle in a follow-up paper. 

\begin{acknowledgments}
The authors would like to thank Adi Nusser for useful discussions, and to Hans A. Winther for kindly providing us with \textsc{mgcamb} version for specific forms of $\mu(a,k)$ and $\gamma(a,k)$ functions implementing our nDGP models. We would also like to thank our anonymous referee for their comments which improved the clarity and robustness of our work. The authors acknowledge the support of the Polish Ministry of Science and Higher Education MNiSW grant DIR/WK/2018/12, as well as the research project ``VErTIGO'' funded by the National Science Center, Poland, under agreement number 2018/30/E/ST9/00698. JEGF is supported by the Spanish Ministry of Universities through a Mar\'ia Zambrano grant with reference UP2021-022, funded within the European Union-Next Generation EU. JEGF acknowledge the IAC facilities and the personnel of the Servicios Inform\'aticos Comunes (SIC) of the IAC. MB is supported by the Polish National Science Center through grants no. 2020/38/E/ST9/00395, 2018/30/E/ST9/00698, 2018/31/G/ST9/03388 and 2020/39/B/ST9/03494

\end{acknowledgments}

\bibliography{v12.bib}

\end{document}